\documentclass[conference]{IEEEtran}

\usepackage{cite}
\usepackage[inline]{enumitem}

\usepackage{cuted}

\usepackage{booktabs} 
\usepackage[numbers,sort&compress]{natbib}

\usepackage{graphicx}
\usepackage[justification=centering]{caption}
\usepackage{wrapfig}
\usepackage{subcaption}

\usepackage{color}
\newcommand{\redHL}[1]{\textcolor{red}{#1}}

\usepackage{amsmath}

\usepackage{multirow}
\begin{document}

\bstctlcite{IEEEexample:BSTcontrol}

\title{Efficient and Scalable MIV-transistor with Extended Gate in Monolithic 3D Integration

(Accepted in MWSCAS 2023)}

\author{ \IEEEauthorblockN{Madhava Sarma Vemuri} \IEEEauthorblockA{ Department of Electrical and Computer Engineering\\
  North Dakota State University\\
  Fargo, North Dakota 58105 \\
  }
\and
\IEEEauthorblockN{Umamaheswara Rao Tida} \IEEEauthorblockA{ Department of Electrical and Computer Engineering\\
  North Dakota State University\\
  Fargo, North Dakota 58105 \\
  }
\thanks{Madhava Sarma Vemuri is with the Department of Electrical and Computer Engineering, North Dakota State University, North Dakota, United States -- 58102}
\thanks{Umamaheswara Rao Tida is with the Department of Electrical and Computer Engineering, North Dakota State University, North Dakota, United States -- 58102. e-mail: umamaheswara.tida@ndsu.edu}
\thanks{Manuscript received: Oct 14, 2019;}
\vspace{-1cm}}

{}


\maketitle

\begin{abstract}
Monolithic 3D integration has become a promising solution for future computing needs. The metal inter-layer via (MIV) forms interconnects between substrate layers in Monolithic 3D integration. Despite small size of MIV, the area overhead can become a major limitation for efficient M3D integration and, thus needs to be addressed. Previous works focused on the utilization of the substrate area around MIV to reduce this area overhead significantly but suffers from increased leakage and scaling factors. In this work, we discuss MIV-transistor realization that addresses both leakage and scaling issue along with similar area overhead reduction compared with previous works and, thus can be utilized efficiently. Our simulation results suggest that the leakage current $(I_{D,leak})$ has reduced by $14K\times$  and, the maximum current $(I_{D,max})$ increased by $58\%$ for the proposed MIV-transistor compared with the previous implementation. In addition, performance metrics of the  inverter realization with our proposed MIV-transistor specifically the delay, slew time and power consumption  reduced by $11.6\%$,  $17.9\%$ and, $4.5\%$ respectively compared with the previous implementation with same MIV area overhead reduction.

\end{abstract}

\begin{IEEEkeywords}
 Monolothic 3D ICs, vertical integration, on-chip devices
\end{IEEEkeywords}

\IEEEpeerreviewmaketitle

\section{Introduction}\label{sec:Intro}

Monolithic three-dimensional integrated circuits (M3D-IC) are realized by sequential integration of ultra-thin substrate layers (thickness of 7nm - 100nm) at low temperatures i.e., below  $500^o C$ \cite{jiang2019ultimate}. The thermal constraint of below  $500^o C$ is levied on the M3D-IC process to ensure the stability of bottom-layer devices \cite{batude_low_temp,batude_low_temp2,mosfet_thermal_stability}. Recent demonstrations such as CoolCube\textsuperscript{TM} and Hydrogen Ion-Cut process show that the M3D-IC process integration is feasible under this thermal constraint \cite{coolcube_batude,low_temp_ion_cut, batude_low_temp2}. 


Metal Inter-layer Via (MIV) provides interconnects between different layers in M3D-IC technology. Due to thin substrate, the MIV thickness has reduced significantly and is comparable to the standard cell i.e, $< 60nm$ \cite{samal2016monolithic, liu2012design}. This reduction in MIV size compared to through-silicon-via (TSV) size i.e, $>2 \mu m$ in conventional 3D stacking has resulted in finer integration of M3D-IC with improved overall alignment accuracy, increased via density and reduced the interconnect routing \cite{tida2014efficacyTLVSI, tida2014novelJETC}. For 14nm design rules, it is predicted that using M3D integration techniques, MIV density over 100 million/mm\textsuperscript{2} can be achieved\cite{bataude_2014_MIV_desnity}. But increase in MIV density reduces the footprint scaling since MIV occupies the substrate region \cite{liu2012design}. In addition, coupling between MIV and devices around it impact the performance of the near by transistors significantly and hence a Keep-out-zone (KOZ) around MIV is required \cite{vemuri_isqed}. Therefore, this area overhead by MIV on the substrate layer should be addressed for efficient M3D-IC implementation.


Previous works have focused on reducing this area overhead by utilizing the substrate region around MIV to form MIV-devices such as MIV-capacitor and MIV-transistor where the MIV serves two purposes i.e., as an interconnect and the device terminal \cite{madhava2020MWSCAS,tida2020SOCC}. With utilizing MIV-transistor for basic inverter circuit, an area savings of 24\% is achieved. However, the MIV-transistor implementation in previous works suffers from scalability issue since the width of the transistor cannot be increased significantly and the higher leakage current due to limited channel control. In this paper, we present the extended gate MIV-transistor to improve the channel control for reducing leakage and addressing scalability since the transistor can be designed for a given width specification. The major contributions of this work are as follows:
\begin{enumerate}
    \item An extended gate dual purpose MIV-transistor is proposed, and its device characteristics are discussed in detail.
    \item We demonstrated the efficacy of the extended gate MIV-transistor in M3D-IC process by comparing with the previous works. 
    \item We studied Inverter models in M3D-IC process using the MIV-based transistors and, extended the study by comparing Ring oscillator using 3-stage inverter chain.
\end{enumerate}

The remaining organization of the paper is as follows: Section \ref{sec:mivfet_goc} discusses the structure of proposed MIV-transistor model and its characteristics. Section \ref{sec:simulation_results} compares  the transistor-level implementation of Inverter and 3-Stage ring oscillator with the previous MIV-transistor model. Finally, the concluding remarks are given in Section \ref{sec:conclusion}.


\section{ Characteristics of Extended Gate MIV-transistor model}\label{sec:mivfet_goc}


 
 \begin{figure}[htp] 
  \centering
 \begin{subfigure}[b]{0.49\linewidth}
\includegraphics[width=\linewidth]{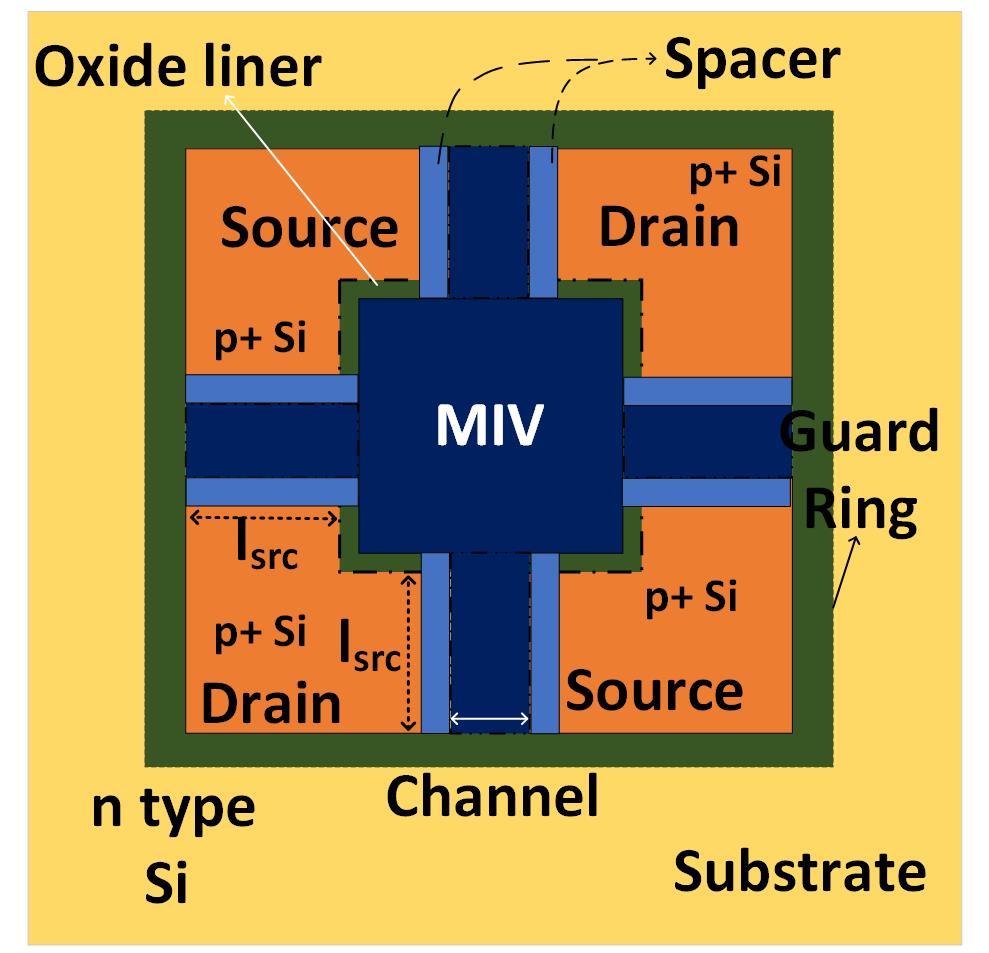}
\caption{Top view}\label{subfig:goc_top_view}
\end{subfigure}
\begin{subfigure}[b]{0.49\linewidth}
\includegraphics[width=\linewidth]{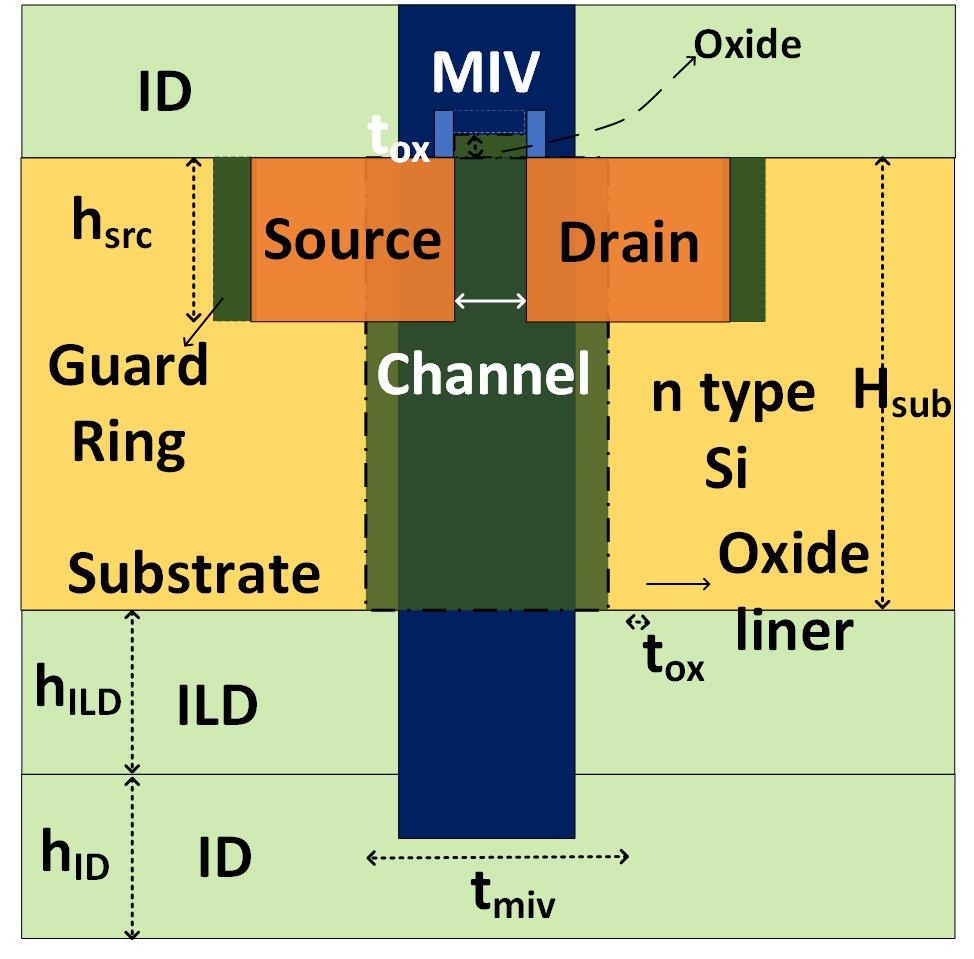}
\caption{Side view}\label{subfig:goc_side_view}
\end{subfigure}
\caption{MIV-transistor model (not to scale)} \label{fig:Transistor_model}
\end{figure}

In this work, MIV-transistor models and circuits are created using Sentaurus TCAD tool. The structure of extended gate MIV-transistor model is shown in Figure \ref{fig:Transistor_model}. The transistor model implemented in \cite{madhava2020MWSCAS} has not assumed the gate extension across the channel region around MIV and, thus suffers from scalability and increased leakage. This extended gate MIV-transistor model has improved channel control and can be scalable thus addressing the issues of previous model. The assumed process parameters are presented in Table \ref{tab:transistor_spec}. The MIV and interconnect regions are modeled using Copper (Cu). The substrate region is created using Silicon (Si). The p-type and n-type regions are modeled using Boron (B) and Arsenic (As). The MIV liner and oxide regions such as Interlayer Dielectric (ILD) and Interconnect Dielectric (ID) are created using Silicon Dioxide material (SiO\textsubscript{2}). The Silicon Nitride (Si\textsubscript{3}N\textsubscript{4}) is used as spacer material. For simulations, we have used  Shockley–Read–Hall (SRH) recombination model along with Poisson equations to model the carrier behavior in TCAD. The active region (n\textsubscript{src}) and substrate region (n\textsubscript{sub}) is doped using a carrier concentration $1\times10^{19}$cm\textsuperscript{-3} and $1\times10^{17}$cm\textsuperscript{-3} respectively. The thickness of MIV is assumed to be $25$nm. The height of substrate is assumed to be $50$nm. We have assumed the channel length to be $14$nm. The gate oxide and liner thickness is assumed to be $1$nm. The nominal values of length and height of active region (l\textsubscript{src} and h\textsubscript{src}) are assumed to be $32$nm and $7$nm respectively. 

\begin{figure}
  \centering
  \begin{subfigure}[b]{0.47\linewidth}
\includegraphics[width=\linewidth]{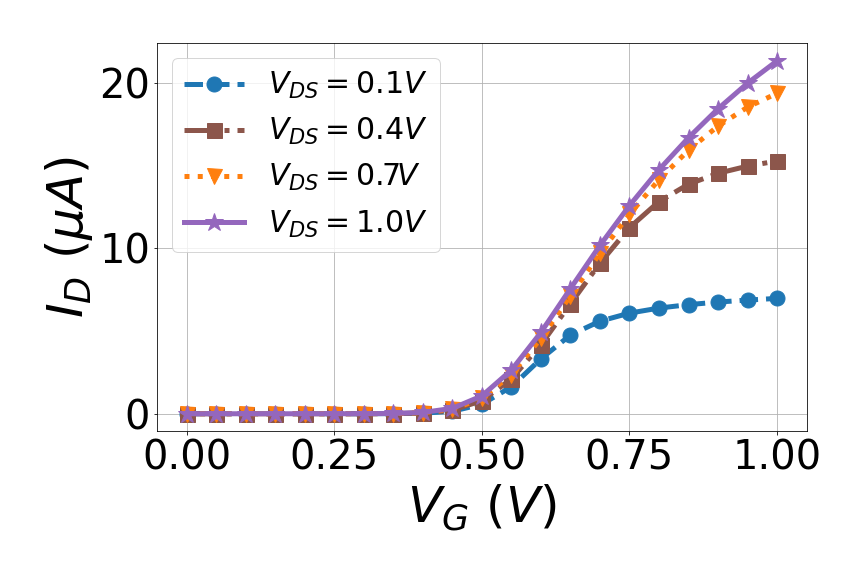}
\caption{$I_D$ v.s. $V_{GS}$ $n-type$}\label{subfig:IdVgs_n}
\end{subfigure}
\begin{subfigure}[b]{0.47\linewidth}
\includegraphics[width=\linewidth]{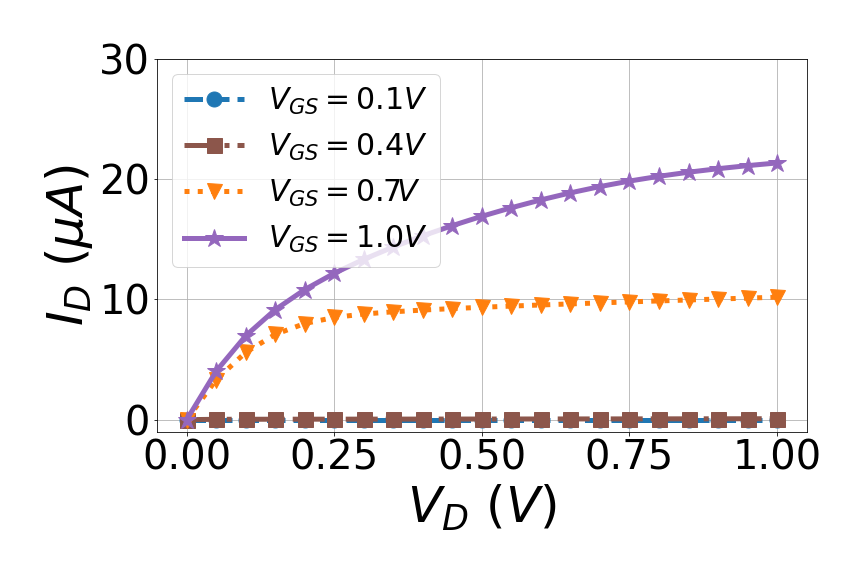}
\caption{$I_D$ v.s. $V_{DS}$ $n-type$}\label{subfig:IdVds_n}
\end{subfigure}
  \begin{subfigure}[b]{0.47\linewidth}
\includegraphics[width=\linewidth]{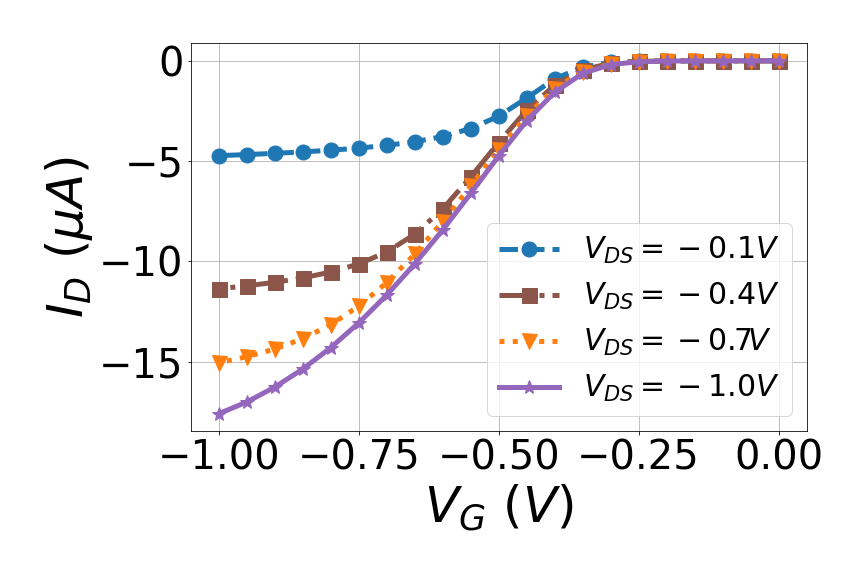}
\caption{$I_D$ v.s. $V_{GS}$ $p-type$}\label{subfig:IdVgs_p}
\end{subfigure}
\begin{subfigure}[b]{0.47\linewidth}
\includegraphics[width=\linewidth]{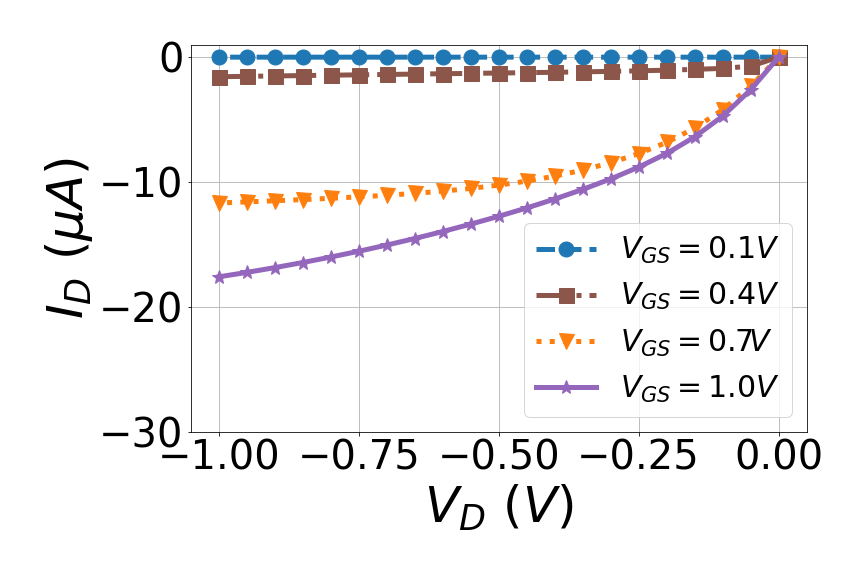}
\caption{$I_D$ v.s. $V_{DS}$ $p-type$}\label{subfig:IdVds_p}
\end{subfigure}
\caption{Characteristics of MIV-transistor}\label{fig:tran_nmos}
\end{figure}

To understand the model behavior, we have simulated the MIV-transistor characteristics ($I_{D}$ v.s. $V_{GS}$ \& $I_{D}$ v.s. $V_{DS}$) for n-type and p-type models in Figure \ref{fig:tran_nmos}. The $I_{D}$ v.s. $V_{DS}$ characteristics for n-type and p-type are presented in Figures \ref{subfig:IdVds_n} and \ref{subfig:IdVds_p} respectively.The $I_{D}$ v.s. $V_{GS}$ characteristics for n-type and p-type are presented in Figures \ref{subfig:IdVgs_n} and \ref{subfig:IdVgs_p} respectively. From the simulations, the threshold voltage ($V_{th}$) is measured to be  $0.51V$ and $-0.37V$ respectively. The maximum drain current ($I_{D,max}$ is $I_{D}$ when $|V_{GS}|=1V$ and $|V_{DS}|=1V$) is measured to be $21.34\mu A$ and $-17.60\mu A$ respectively. The leakage current ($I_{D,leak}$ is $I_{D}$ when $|V_{GS}|=0V$ and $|V_{DS}|=1V$) is measured to be $0.46pA$ and $-21.71pA$ respectively.


\begin{table}
\caption{Process and design parameters of MIV-transistor}
\label{tab:transistor_spec}
\begin{center}
\begin{tabular}{|c|c|p{3.5cm}|c|}
    \hline
    & \textbf{Parameter} &  \textbf{Description} & \textbf{Value}\\
    \hline
    \multirow{5}{*}{\rotatebox{90}{\textbf{Process}}}& 
    H\textsubscript{sub} & Height of substrate  & $50$ nm \\
    \cline{2-4}
    & h\textsubscript{src} & Height of source/drain region & $7$ nm \\
    \cline{2-4}
    & t\textsubscript{ox} & Thickness of oxide liner & $1$ nm \\
    \cline{2-4}
    & n\textsubscript{sub} & Substrate doping & $1\times10^{17}$cm\textsuperscript{-3}\\
    \cline{2-4}
    & n\textsubscript{src} & Source/Drain doping & $1\times10^{19}$cm\textsuperscript{-3} \\
    \hline
    \multirow{3}{*}{\rotatebox{90}{\textbf{Design}}}& 
    t\textsubscript{miv} & MIV thickness & $25$ nm \\
    \cline{2-4}
    & l\textsubscript{src} & Length of source/drain region & $32$ nm \\
    \cline{2-4}
    & l\textsubscript{channel} & Length of channel & $14$ nm \\
    \hline
\end{tabular}
\end{center}

\end{table}

\section{Simulation Results}\label{sec:simulation_results}

\begin{figure}[htp]
    \centering
    \includegraphics[width=0.8\linewidth]{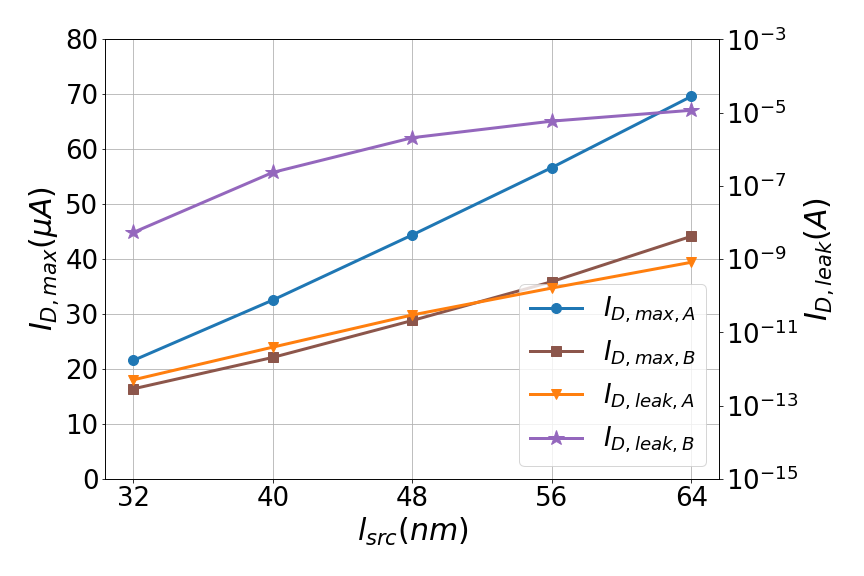}
    \caption{$I_{D,max}$ and $I_{D,leak}$ v.s. $l_{src}$}
    \label{fig:varying_lsrc}
\end{figure}

In this section, we compare the extended gate MIV-transistor models with the previous model presented in \cite{madhava2020MWSCAS, tida2020SOCC} for M3D-Inverter. For simplicity, we only compare the n-type MIV transistor models.
In the previous models, the length of source/drain region $l_{src}$ of MIV-transistor is limited by the extent of inversion region created by the MIV based gate region. As  $l_{src}$ increases, the extent of control by the gate reduced resulting in increased leakage currents ($I_{D,leak}$) as shown in Figure \ref{fig:varying_lsrc}. In the proposed MIV-transistor model (denoted by A and previous model denoted by B), this effect is alleviated due to the presence of gate over channel regions of MIV-transistor. As the $l_{src}$ is varied form $32$nm to $64$nm, maximum current ($I_{D,max}$) scaled linearly for both the models. The $I_{D,max}$ increased by 58\%  when active region length was at maximum ($l_{src}=64nm$). For the same case, $I_{D,leak}$ reduced by $14K\times$ in the proposed model resulting in a significant improvement in the MIV-transistor characteristics and, hence the extended gate MIV-transistor model can be scaled depending on the specifications with reasonable leakage current.

\begin{figure*}[htp]
  \centering
    \begin{subfigure}[b]{0.25\linewidth}
   \centering
\includegraphics[width=\linewidth]{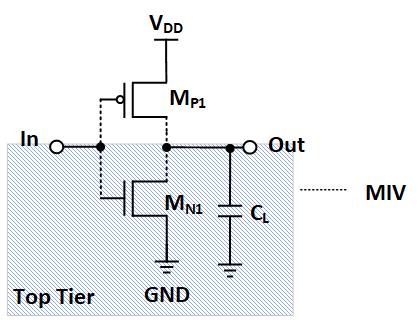}
\caption{Schematic}\label{subfig:inv_circuit}
\end{subfigure}
  \begin{subfigure}[b]{0.35\linewidth}
\includegraphics[width=\linewidth]{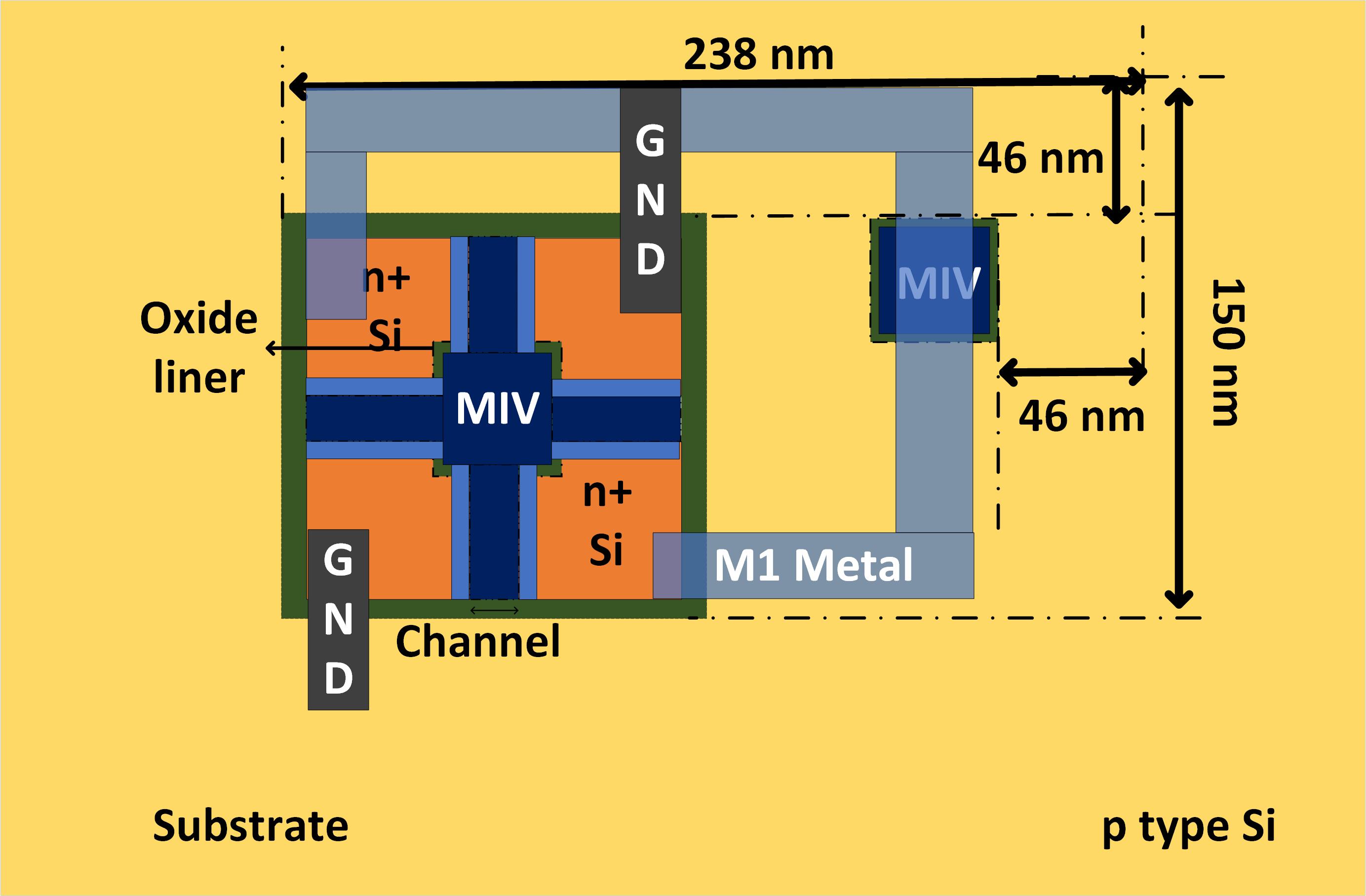}
\caption{Top View}\label{subfig:inv_goc_top}
\end{subfigure}
\begin{subfigure}[b]{0.35\linewidth}
\includegraphics[width=\linewidth]{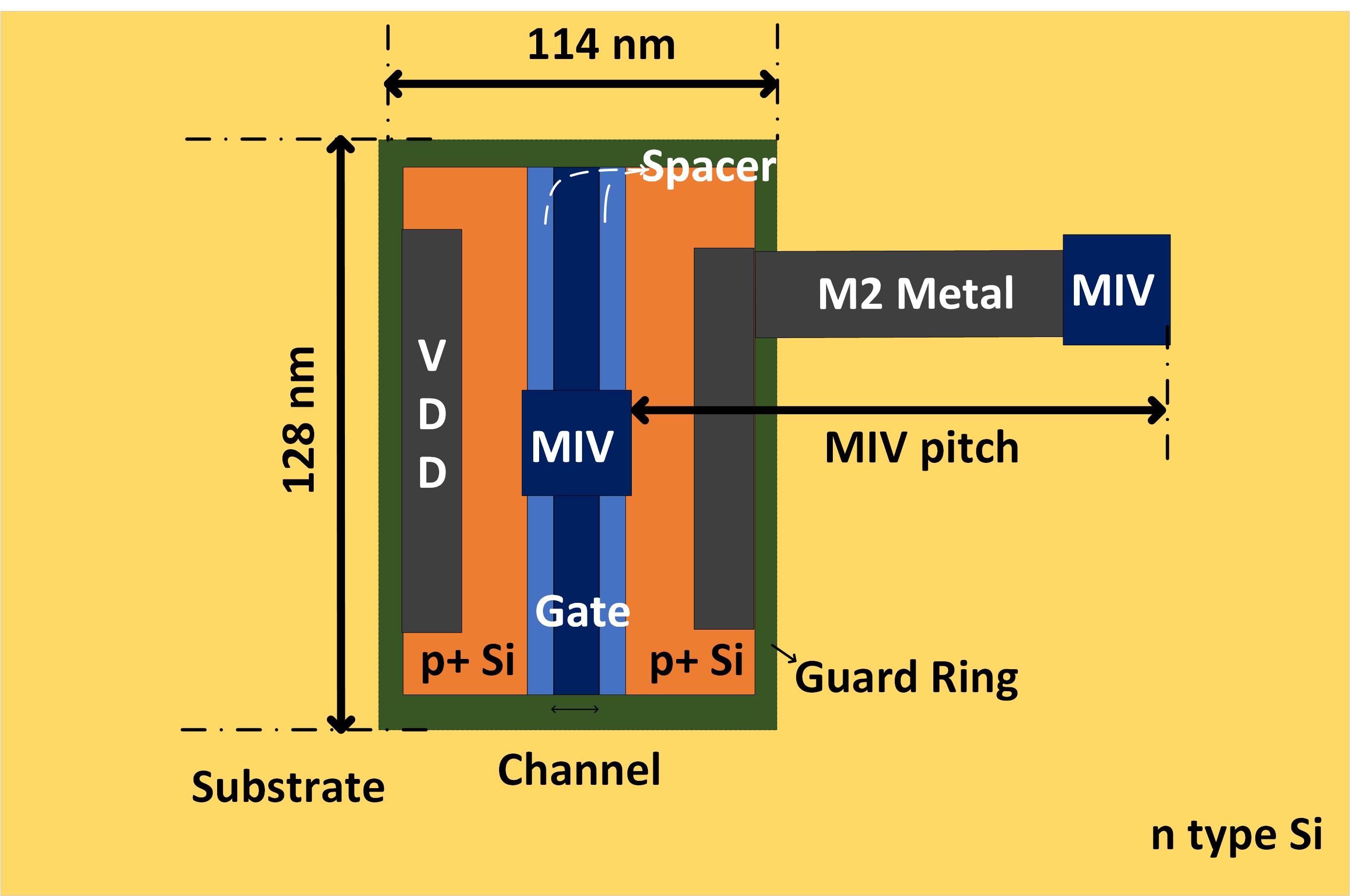}
\caption{Bottom View}\label{subfig:inv_goc_bot}
\end{subfigure}\label{fig:inv_goc_bot}

\caption{M3D-Inverter Layout (not to scale)}\label{fig:inv_goc_layout}
\end{figure*}

\subsection{M3D-Inverter}\label{sec:inv_goc}

In this section, we have created inverter model to compare the gate level performance of MIV-transistor models. We first look into the design of Inverter in M3D-IC and, extended the study for 3-stage Inverter based Ring Oscillator. In the design of M3D-Inverter, we have assumed a 2-layer M3D process where the bottom layer is assigned to p-type transistor (M\textsubscript{P1}) and, the top layer is assigned to n-type transistor (M\textsubscript{N1}) as shown in Figure \ref{subfig:inv_circuit}. We have used 2 Metal (M1 \& M2) routing layers to route the interconnection between active layer and MIV. The process and design parameters for the MIV-transistor implementation is given Table \ref{tab:transistor_spec} and rest of the parameters used for the circuit implementation in two-tier process are given in both Table \ref{tab:m3d_process_spec}. MIV connects the top most routing layer (M2) of the bottom substrate and, passes through the top substrate to connect the bottom routing layer (M1) of the top substrate as shown in figure \ref{subfig:inv_goc_top} and \ref{subfig:inv_goc_bot}. Based on the study of impact on adjacent devices by the MIV presence, we have assumed the keep-out-zone of MIV to be $46nm$, where the leakage increase of adjacent device due to MIV presence is under $10\times$ \cite{vemuri_isqed}. For convenience, we have assumed the M1 and M2 pitch to be $46nm$. The dimension of Interconnect routing layers are based on assumptions from \cite{chang2017impact}. The layout followed a similar construction presented in \cite{madhava2020MWSCAS,tida2020SOCC}, however we have used the proposed MIV-transistor replacing the previous MIV-transistor used for M\textsubscript{N1} transistor. The new model requires no silicon area overhead compared with the previous model since it follows a similar layout dimensions.

\begin{table}
\caption{Additional process and design parameters for M3D-Inverter and M3D-Ring Oscillator designs}
\label{tab:m3d_process_spec}
\begin{center}
\begin{tabular}{|c|p{5cm}|c|}
    \hline
     \textbf{Parameter} &  \textbf{Description} & \textbf{Value}\\
    \hline
    T\textsubscript{gaurd} & Thickness of Guard Ring  & $14$ nm \\
    \hline
    M\textsubscript{x,wid} & Width of Metal Routing layers 1,2  & $20$ nm \\
    \hline
    M\textsubscript{x,th} & Thickness of Metal Routing layers 1,2  & $40$ nm \\
    \hline
    via\textsubscript{contact} & Width of contact  & $18$ nm \\
    \hline
    MIV\textsubscript{pitch} & Minimum separation between MIV  & $73$ nm \\
    \hline
    l\textsubscript{p1} & Length of active region of p-type transistor in bottom layer  & $32$ nm \\
    \hline
    W\textsubscript{p1} & Width of p-type transistor in bottom layer  & $100$ nm \\
    \hline
    l\textsubscript{n1} & Length of active region of n-type MIV-transistor in top layer  & $32$ nm \\
    \hline
    t\textsubscript{spacer} & Thickness of Si\textsubscript{3}N\textsubscript{4} Spacer  & $5$ nm \\
    \hline
\end{tabular}
\end{center}
\end{table}

\begin{figure}[htp] 
  \centering
\includegraphics[width=0.75\linewidth]{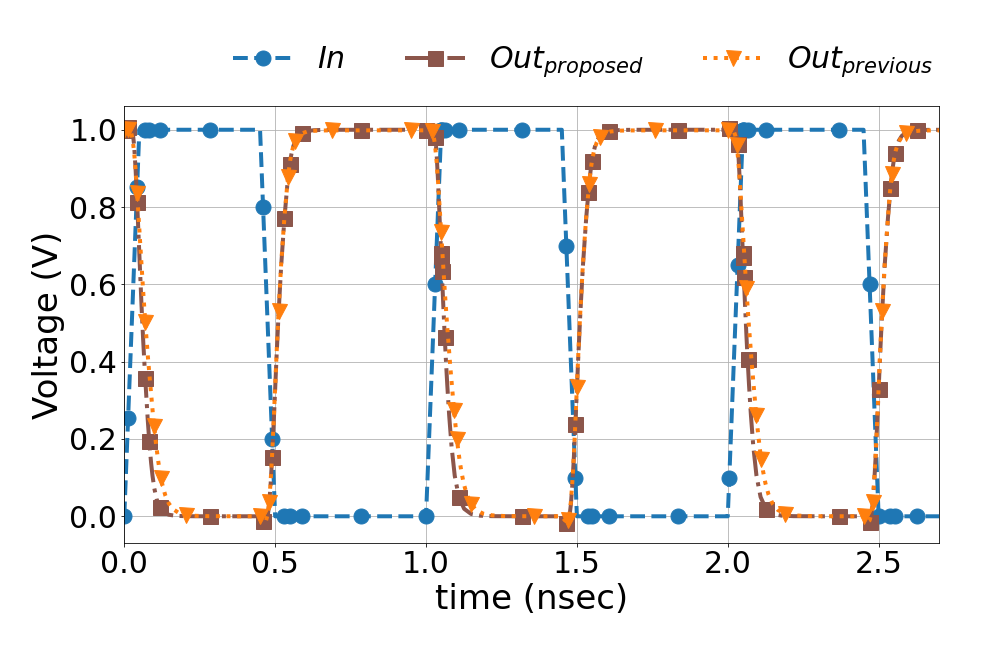}
\caption{M3D-Inverter Simulation Result}\label{subfig:inv_simualtions}
\end{figure}

The transient simulation results for M3D-Inverter using the \begin{enumerate*}[label=\Alph*)]
    \item Proposed model and
    \item Previous model
    \end{enumerate*}
with a capacitive load of 1fF and, input with a time period of 1ns is shown in Figure \ref{subfig:inv_simualtions}. From the simulation results, the fall time $t_{f}$ (time interval to drop the output from 90\% to 10\%) and rise time $t_{r}$ (time interval to raise the output from 10\% to 90\%) are measured to be $60.2$ps and $58.7$ps respectively. The propagation delay from Low to High $t_{pLH}$ (time interval between 50\% input to 50\% output when the output transitions from Low to High) and $t_{pHL}$ (time interval between 50\% input to 50\% output when the output transitions from High to Low) are measured to be $34.9$ps and $34.9$ps respectively. The $t_{r}$ and  $t_{pLH}$ reported no significant difference, and are changed by $-0.5$\% and $1$\% respectively. Since n-type transistor differs in both models, there is a significant change relating to $t_{f}$ and  $t_{pHL}$, and are improved by almost 31\% and 21.4\% respectively. Also, Table \ref{tab:m3div_comparison} gives the average delay $t_{delay}$ (average of $t_{pLH}$ and $t_{pHL}$), slew time $t_{slew}$ (average of $t_{r}$ and $t_{f}$) and total power consumption $Power$. From the table, we can see that  $t_{delay}$ and $t_{slew}$ reduced by 11.6\% and 17.9\% respectively compared with previous MIV-transistor inverter design. Finally, the total power consumption of the inverter design is $1.16\mu W$ with our implementation and is reduced by $4.5$\% compared with previous work. The area occupied by both the models are $0.0503\ \mu m^2$ and the improvements are seen in the proposed model with no significant changes in the silicon area utilization.


\begin{table}[]
\caption{Comparison of 3D-Inverter results}\label{tab:m3div_comparison}
\centering
    \begin{tabular}{|c|c|c|c|}
    \hline
        \textbf{Parameter} & \textbf{Previous} & \textbf{Proposed} & \textbf{diff.\%} \\
        \hline 
        \textbf{$t_{delay}\ (psec)$} & 39.5 & 34.9 & - 11.6\% \\
        \hline         
        \textbf{$t_{slew}\ (psec)$} & 72.5 & 59.5 & - 17.9\% \\
        \hline         
        \textbf{$Power (\mu W)$}	& 1.22 & 1.16 & - 4.5\% \\
        \hline 
    \end{tabular}
\end{table}

\subsection{M3D-Ring Oscillator}\label{sec:inv_3stage}

\begin{figure}[htp] 
  \centering
  \begin{subfigure}[b]{\linewidth}
   \centering
\includegraphics[width=0.75\linewidth]{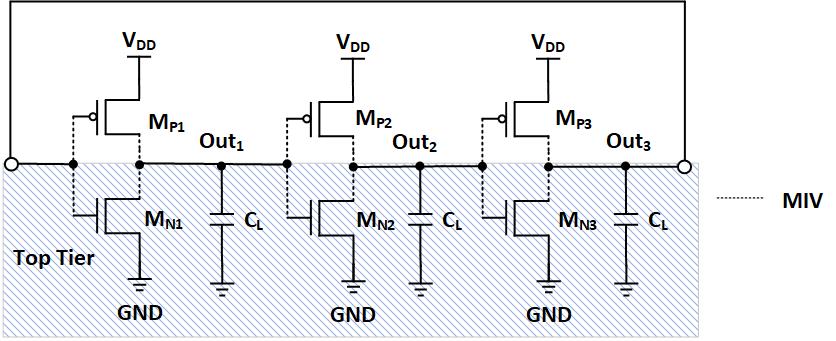}
\caption{Schematic}\label{subfig:3stage_inverter_chain}
\end{subfigure}
\begin{subfigure}[b]{\linewidth}
  \centering
\includegraphics[width=0.7\linewidth]{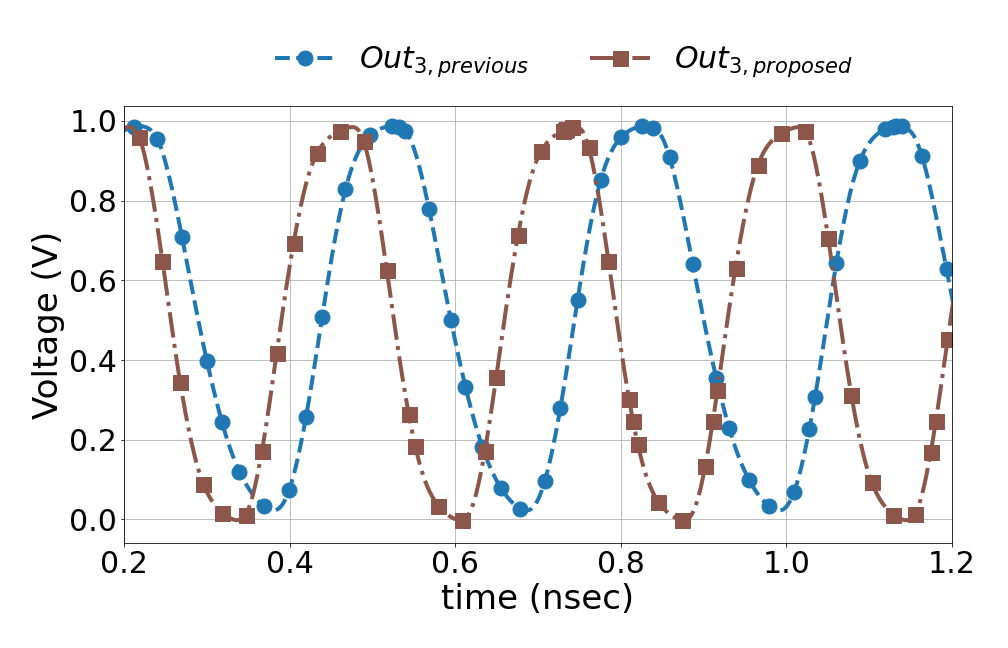}
\caption{Simulation Results}\label{subfig:3stage_inverter_simulation}
\end{subfigure}
\caption{M3D-Ring Oscillator}\label{subfig:3stage_inverter}
\end{figure}

The M3D-Ring Oscillator is created using 3 stage M3D-inverter connected back to back as shown in Figure \ref{subfig:3stage_inverter_chain}. In the schematic, a feedback loop is created by connecting the final output (Out\textsubscript{3}) and is fed back as the input to the  M3D-Ring Oscillator. Each inverter within the M3D-Ring Oscillator followed a similar M3D process presented in Table \ref{tab:transistor_spec} and \ref{tab:m3d_process_spec}.  A capacitive load of $1$fF is placed at the outputs of each M3D-inverter. The simulation results of the  proposed and previous M3D-Ring Oscillator are shown in \ref{subfig:3stage_inverter_simulation}. The frequency of oscillations for extended gate MIV-transistor and previous MIV-transistor based ring oscillator is $3.27GHz$ and $3.71GHz$ respectively. The total Power consumption for  proposed model and the previous model is $13.6\mu W$ and $11.6\mu W$ respectively. The total Energy consumption of the  extended gate MIV-transistor based oscillator is $3.7 pJ$, an increase of $3.1$\% compared with previous MIV-transistor based oscillator (Note: We are comparing total Energy consumption instead of total Power consumption since both models are oscillating at different frequencies).  The area occupied by both the models is $0.151\ \mu m^2$.


\section{Conclusion}\label{sec:conclusion}
This paper discusses the efficacy of MIV-transistor by extending its region of control. Our Simulation results suggest that, the extended gate MIV-transistor has reduction in leakage current by $14K \times$ and improvement in maximum current by $58\%$ compared with previous MIV-transistor implementation. We have also implemented M3D-Inverter and M3D-Ring Oscillator to compare the performance metrics of the MIV-transistor. The simulation results suggest that, with no additional area overhead, the proposed MIV-transistor outperforms the previous models. 


\section*{Acknowledgement}
This work is supported by National Science Foundation under Award number -- 2105164.

\bibliographystyle{IEEEtran}
\bibliography{main}

\begin{thebibliography}{10}
\providecommand{\url}[1]{#1}
\csname url@samestyle\endcsname
\providecommand{\newblock}{\relax}
\providecommand{\bibinfo}[2]{#2}
\providecommand{\BIBentrySTDinterwordspacing}{\spaceskip=0pt\relax}
\providecommand{\BIBentryALTinterwordstretchfactor}{4}
\providecommand{\BIBentryALTinterwordspacing}{\spaceskip=\fontdimen2\font plus
\BIBentryALTinterwordstretchfactor\fontdimen3\font minus
  \fontdimen4\font\relax}
\providecommand{\BIBforeignlanguage}[2]{{%
\expandafter\ifx\csname l@#1\endcsname\relax
\typeout{** WARNING: IEEEtran.bst: No hyphenation pattern has been}%
\typeout{** loaded for the language `#1'. Using the pattern for}%
\typeout{** the default language instead.}%
\else
\language=\csname l@#1\endcsname
\fi
#2}}
\providecommand{\BIBdecl}{\relax}
\BIBdecl

\bibitem{jiang2019ultimate}
J.~Jiang, K.~Parto \emph{et~al.}, ``{Ultimate Monolithic-3D Integration With 2D
  Materials: Rationale, Prospects, and Challenges},'' \emph{IEEE Journal of the
  Electron Devices Society}, 2019.

\bibitem{batude_low_temp}
P.~Batude, L.~Brunet \emph{et~al.}, ``{3D Sequential Integration:
  Application-driven technological achievements and guidelines},'' in
  \emph{IEEE International Electron Devices Meeting (IEDM)}, 2017.

\bibitem{batude_low_temp2}
C.~Fenouillet-Beranger, B.~Previtali \emph{et~al.}, ``{FDSOI bottom MOSFETs
  stability versus top transistor thermal budget featuring 3D monolithic
  integration},'' in \emph{44th European Solid State Device Research Conference
  (ESSDERC)}, 2014.

\bibitem{mosfet_thermal_stability}
------, ``{FDSOI bottom MOSFETs stability versus top transistor thermal budget
  featuring 3D monolithic integration},'' in \emph{44th European Solid State
  Device Research Conference (ESSDERC)}, 2014.

\bibitem{coolcube_batude}
P.~Batude, C.~Fenouillet-Beranger \emph{et~al.}, ``{3DVLSI with CoolCube
  process: An alternative path to scaling},'' in \emph{Symposium on VLSI
  Technology (VLSI Technology)}.\hskip 1em plus 0.5em minus 0.4em\relax IEEE,
  2015.

\bibitem{low_temp_ion_cut}
H.~Han, R.~Choi \emph{et~al.}, ``Low temperature and ion-cut based monolithic
  3d process integration platform incorporated with cmos, rram and photo-sensor
  circuits,'' in \emph{IEEE International Electron Devices Meeting
  (IEDM)}.\hskip 1em plus 0.5em minus 0.4em\relax IEEE, 2020.

\bibitem{samal2016monolithic}
S.~K. Samal, D.~Nayak \emph{et~al.}, ``{Monolithic 3D IC vs. TSV-based 3D IC in
  14nm FinFET technology},'' in \emph{IEEE SOI-3D-Subthreshold Microelectronics
  Technology Unified Conference (S3S)}.\hskip 1em plus 0.5em minus 0.4em\relax
  IEEE, 2016.

\bibitem{liu2012design}
C.~Liu and S.~K. Lim, ``{A design tradeoff study with monolithic 3D
  integration},'' in \emph{Thirteenth International Symposium on Quality
  Electronic Design (ISQED)}.\hskip 1em plus 0.5em minus 0.4em\relax IEEE,
  2012.

\bibitem{tida2014efficacyTLVSI}
U.~R. Tida, R.~Yang \emph{et~al.}, ``{On the efficacy of through-silicon-via
  inductors},'' \emph{IEEE Transactions on Very Large Scale Integration (VLSI)
  Systems}, 2014.

\bibitem{tida2014novelJETC}
U.~R. Tida, C.~Zhuo, and Y.~Shi, ``{Novel through-silicon-via inductor-based
  on-chip DC-DC converter designs in 3D ICs},'' \emph{ACM Journal on Emerging
  Technologies in Computing Systems (JETC)}, 2014.

\bibitem{bataude_2014_MIV_desnity}
P.~Batude, B.~Sklenard \emph{et~al.}, ``{3D sequential integration
  opportunities and technology optimization},'' in \emph{IEEE International
  Interconnect Technology Conference}, 2014.

\bibitem{vemuri_isqed}
M.~S. Vemuri and U.~Rao~Tida, ``{Metal Inter-layer Via Keep-out-zone in M3D IC:
  A Critical Process-aware Design Consideration},'' in \emph{24th International
  Symposium on Quality Electronic Design (ISQED)}, 2023.

\bibitem{madhava2020MWSCAS}
------, ``{Dual-Purpose Metal Inter-layer Via Utilization in Monolithic
  Three-Dimensional (M3D) Integration},'' in \emph{IEEE 63rd International
  Midwest Symposium on Circuits and Systems (MWSCAS)}, 2020.

\bibitem{tida2020SOCC}
U.~R. Tida and M.~S. Vemuri, ``{Efficient Metal Inter-Layer Via Utilization
  Strategies for Three-dimensional Integrated Circuits},'' in \emph{IEEE 33rd
  International System-on-Chip Conference (SOCC)}, 2020.

\bibitem{chang2017impact}
K.~Chang, K.~Acharya \emph{et~al.}, ``{Impact and design guideline of
  monolithic 3-D IC at the 7-nm technology node},'' \emph{IEEE Transactions on
  Very Large Scale Integration (VLSI) Systems}, 2017.

\end{thebibliography}

\end{document}